\documentclass[prb,aps,preprint,showpacs,preprintnumbers,amsmath,amssymb,superscriptaddress]{revtex4}

\usepackage{epsfig}
\usepackage{graphicx}
\usepackage{dcolumn}
\usepackage{bm}

\begin{document}

\title{Construction and solution of a Wannier-functions based Hamiltonian 
in the pseudopotential plane-wave framework 
for strongly correlated materials}

\author{Dm.~Korotin}
\author{A.~V.~Kozhevnikov}
\affiliation{Institute of Metal Physics, Russian Academy of Sciences-Ural 
Division, 620219 Yekaterinburg GSP-170, Russia}
\author{S.~L.~Skornyakov}
\affiliation{Ural State Technical University-UPI,
620002 Yekaterinburg, Russia}
\author{I.~Leonov}
\affiliation{Abdus Salam International Center for Theoretical 
Physics, Strada Costiera 11, 34014 Trieste, Italy}
\author{N.~Binggeli}
\affiliation{Abdus Salam International Center for Theoretical 
Physics, Strada Costiera 11, 34014 Trieste, Italy}  
\affiliation{DEMOCRITOS National Simulation Center, INFM-CNR, 
Trieste, Italy}
\author{V.~I.~Anisimov}
\affiliation{Institute of Metal Physics, Russian Academy of Sciences-Ural 
Division, 620219 Yekaterinburg GSP-170, Russia}
\author{G.~Trimarchi\footnote{now at: National Renewable Energy Laboratory, Golden, CO 80401, USA}}
\affiliation{Abdus Salam International Center for Theoretical Physics, 
Strada Costiera 11, 34014 Trieste, Italy}  
\affiliation{DEMOCRITOS National Simulation Center, INFM-CNR, 
Trieste, Italy}

\date{\today}

\begin{abstract}
\textit{Ab initio} determination of model Hamiltonian parameters for strongly 
correlated materials is a key issue in applying many-particle theoretical tools 
to real narrow-band materials. We propose a self-contained calculation scheme 
to construct, with an {\it ab initio} approach, and solve such a Hamiltonian. The 
scheme uses a Wannier-function-basis set, with the Coulomb interaction parameter $U$ 
obtained specifically for these Wannier functions via constrained Density 
functional theory (DFT) calculations. The Hamiltonian is solved by Dynamical 
Mean-Field Theory (DMFT) with the effective impurity problem treated by the 
Quantum Monte Carlo (QMC) method. Our scheme is based on the pseudopotential 
plane-wave method, which makes it suitable for developments addressing the 
challenging problem of crystal structural relaxations and transformations due 
to correlation effects. We have applied our scheme to the ``charge transfer 
insulator'' material nickel oxide and demonstrate a good agreement with the 
experimental photoemission spectra.
\end{abstract}


\maketitle

\section{Introduction}
The problem of inter-electron Coulomb interaction in solids is a many-particle 
problem which cannot be solved without major approximations.  
The most successful and widely used approach is electronic density functional 
theory (DFT) \cite{Kohn-Sham,jones} ---within the local density approximation 
(LDA) or generalized gradient approximation (GGA), where all electrons feel 
the same one-particle potential defined by the electronic density distribution 
in the system. For wide-band materials, where the kinetic energy term dominates 
and the inter-electron Coulomb interaction can be treated in an average way with 
an energy (time) independent potential, this approach works very well. However 
for narrow-band materials this is not the case. The Coulomb interaction dominates 
over the kinetic energy, leading to strong correlations between electrons. Hence 
a static one-electron potential, as in DFT, becomes a bad approximation. 
Nevertheless, DFT can still give good results in such systems for some static 
integral properties. Moreover, DFT calculations can serve as a starting point 
for more sophisticated approaches designed to treat strongly correlated systems.

The physics of strongly correlated systems was historically studied via 
solution of model Hamiltonians, such as the Hubbard \cite{Hubbard} and 
Anderson models \cite{And}. In such Hamiltonians, the Coulomb interaction 
term is defined using a set of localized atomic-like orbitals centered on 
atomic sites. While the kinetic energy is invariant with the choice of the 
wave-functions-basis set of the model, the Coulomb interaction term 
significantly depends on the specific form of the atomic-like orbitals. 

Wannier functions \cite{Wannier} are defined as Fourier transforms of 
Bloch functions, from wavevector to real space. They are considered 
nowaday as an optimal choice of basis set to construct model Hamiltonians,
\cite{Wannier-DMFT1,Wannier-DMFT2} because they have a form of atomic 
centered localized orbitals and represent a complete basis set for the 
Bloch functions Hilbert space. However, Wannier functions are not uniquely 
defined and one needs to impose some additional conditions to make them unique. 

Another source of uncertainty in the construction of the Coulomb interaction 
term is the value of the onsite Coulomb repulsion parameter $U$. Attempts to 
simply determine it via the integral of the Coulomb potential multiplied by 
the squares of the orbital wave functions \cite{Solovyev,Ferdi} gave 
values 2-3 times larger than the experimental estimates. This is due to the 
neglect of strong relaxation and screening effects. The latter effects can 
be calculated using perturbation theory, but the results strongly depend on 
the choice of the screening channels and the number of higher and lower lying 
states included in the expansion series. 

DFT can provide not only good data for the kinetic energy terms in the model 
Hamiltonians, but also gives a practical alternative way to calculate the 
Coulomb parameter $U$ using constrained DFT 
calculations\cite{AG,ConstrainPickett,Coco,nakamura}. In the constrained calculations, 
the DFT equations are solved with a fixed occupancy of the localized orbitals, 
and $U$ is defined as a derivative of the orbital energy with respect to its occupancy.

After its construction, the Hamiltonian for strongly correlated electrons needs 
to be solved in a way that should be as close to exact as possible. Recently Dynamical 
Mean-Field Theory (DMFT)~\cite{vollha93,pruschke,georges96,kotliar} became a very popular tool 
to describe strongly correlated materials, especially when the numerically exact 
Quantum Monte Carlo (QMC)\cite{Hirsh} method is used as a solver for the effective 
impurity problem. While neglecting, in its standard version, the inter-site 
correlations, DMFT fully accounts for the local dynamics. When the system is 
not very close to an ordering-disordering transition, single-site DMFT usually 
provides a satisfactory agreement between calculated and experimental
spectra.

Various methods are available to carry out DFT electronic structure calculations, 
which are based on different approximations for the wave-functions expansion.  
Any one of those methods may be used, in principle, to construct and solve 
the Hamiltonian for strongly correlated materials, if only the electronic 
spectral functions are needed as the results of the calculations. However, 
Coulomb correlation effects can lead to strong renormalization of the 
electron-lattice coupling and hence to complicated phase transitions due 
to lattice distortions. To describe such effects, lattice relaxation should 
be taken into account explicitly. 

The pseudopotential plane-wave method for first-principles DFT calculations 
is well suited and widely used to determine lattice relaxations in 
solids\cite{pseudo} -- in recent years this has also been extended to 
DFT plus onsite Coulomb interaction (DFT + $U$) calculations.\cite{Coco,Fabris05}  
It would be desirable to use this method also as a basis to develop a self-contained 
scheme to determine, via  
DMFT and Wannier functions, the properties of strongly correlated materials. 
So far, DMFT computations with Wannier functions have been implemented either 
using Muffin-Tin-Orbital (LMTO) methods, with Wannier functions constructed 
using the N-order 
muffin-tin-orbital scheme,\cite{Pavarini04} or more recently using some 
mixed-basis methods, with maximally localized Wannier functions,\cite{Wannier-DMFT2} 
and some an adjustable $U$ parameter value. 

In the present work, we propose a calculation scheme where all the terms of 
the Hamiltonian are generated, in a consistent way, within the {\it ab initio} 
pseudopotential plane-wave framework. Starting from the DFT 
pseudopotential-plane-waves method, we construct atomic-centered Wannier 
functions and produce the Hamiltonian kinetic-energy term in the basis of 
the Wannier functions. Then the value of the Coulomb parameter $U$ for 
electrons in these Wannier functions is calculated via constrained DFT 
calculations. At the last stage, the DMFT-QMC method is used to solve 
the Hamiltonian. 
Recently, the electronic structure of nickel oxide was successfully described within the DMFT method\cite{Kunes07}. In the current article NiO plays the role of a well known and already well described test system. The goal is to demonstrate the ability of our new method to reproduce the electronic structure of real system with a level of agreement comparable to previous LMTO-based DMFT calculations. Calculated spectral functions are compared with recent NiO DMFT-calculation
results\cite{Kunes07} and with the experimental photoemission spectra.

\section{Method}
\label{Method} 
\subsection{Wannier functions}
Wannier functions (WFs) $| W_n^{\bf T} \rangle$ are defined as Fourier transforms 
of Bloch functions $| \Psi_{n{\bf k}} \rangle$: \cite{Wannier} 
\begin{equation}
\label{Wannier:Definition}
| W_n^{\bf T} \rangle = \frac{1}{\sqrt{\Omega}} \sum_{\bf k} e^{-i{\bf kT}}| \Psi_{n{\bf k}} \rangle,
\end{equation}
where ${\bf T}$ is the lattice translation vector, $n$ the band number and ${\bf k}$ 
the reciprocal lattice vector. 
WFs are not uniquely defined because, in the single band case, there is a freedom 
of choice of the phases of the Bloch functions, $| \Psi_{n{\bf k}} \rangle$, as a function 
of ${\bf k}$, and in the multiband case, any set of orthogonal linear combination of Bloch 
functions $| \Psi_{n{\bf k}} \rangle$  could be used in (\ref{Wannier:Definition}). 
The uncertainty in the WF's definition corresponds to a 
freedom of choice for a unitary transformation matrix $U^{({\bf k})}_{jn}$
\begin{equation}
\label{Wannier:Umatrix}
|\Psi_{n{\bf k}} \rangle \to \sum_j U^{({\bf k})}_{jn} |\Psi_{j{\bf k}} \rangle.
\end{equation}
One of the most commonly used approach to generate WFs was proposed by N.~Marzari and 
D.~Vanderbilt \cite{MarzariVanderbilt}. They use a condition of maximum localization 
of the WFs, that results in a variational procedure for the matrix $U^{({\bf k})}_{jn}$. 
As an initial step before the variational process, a set of trial localized orbitals 
in the form of atomic orbitals was chosen and projected onto the subspace of Bloch 
functions. Later \cite{KuRosner} it was shown that this initial guess for the WFs 
of transition-metal oxides is usually so good that the variational procedure can be 
dropped and the projection of the trial orbitals onto the subspace of Bloch functions 
can be used to define the unitary transformation matrix $U^{({\bf k})}_{jn}$.

In the present work, we employ the pseudopotential method and a plane wave 
basis set. Hence, site centered pseudoatomic orbitals $ \phi_n$ were chosen 
as a set of trial orbitals. 

Nonorthogonalized approximations to the WFs in the direct $|\widetilde{W}^{\bf T}_n\rangle$ 
and reciprocal space  $|\widetilde{W}_{n{\bf k}}\rangle$ are calculated as projection of the pseudoatomic orbitals onto a subspace of Bloch functions that is defined by setting an energy interval $E_1\le\varepsilon_i({\bf k})\le E_2$ or some band numbers $N_1\le i\le N_2$:
\begin{equation}
|\widetilde{W}^{\bf T}_n\rangle = \sum_{\bf k} |\widetilde{W}_{n{\bf k}}\rangle e^{-i{\bf kT}},
\end{equation}
\begin{equation}
\label{Wannier:proj}
|\widetilde{W}_{n{\bf k}}\rangle \equiv \sum_{i=N_1}^{N_2} |\Psi_{i{\bf k}}\rangle\langle \Psi_{i{\bf k}} | \phi_{n{\bf k}} \rangle = 
\sum_{E_1\le\varepsilon_i({\bf k})\le E_2} |\Psi_{i{\bf k}}\rangle\langle \Psi_{i{\bf k}} | \phi_{n{\bf k}} \rangle.
\end{equation}

In the plane waves basis, Bloch functions and Bloch sums of pseudoatomic orbitals $|\phi_{n{\bf k}}\rangle = \sum_{\bf T} e^{i{\bf kT}}|\phi_n^{\bf T}\rangle $  can be decomposed as:
\begin{equation}
|\Psi_{i{\bf k}} \rangle = \sum_{\bf q} c_{i,{\bf q}} ({\bf k}) | {\bf k-q} \rangle,
\label{Methods:BF}
\end{equation}
\begin{equation}
|\phi_{n{\bf k}} \rangle = \sum_{\bf q^{\prime}} a_{n,{\bf q^{\prime}}} ({\bf k}) | {\bf k-q^{\prime}} \rangle,
\end{equation}
where $n$ is a combined index $jlm\sigma$ ($j$ is the index for the atom, $lm$ are the orbital and magnetic quantum numbers, respectively, and $\sigma$ is the spin projection).

Using Eq. (\ref{Wannier:proj}), together with these decompositions,  one obtains:
\begin{eqnarray}
\label{WF-defin}
|\widetilde{W}_{n{\bf k}}\rangle & \equiv \sum \limits_{i=N_1}^{N_2} |\Psi_{i{\bf k}}\rangle\langle \Psi_{i{\bf k}} | \phi_{n{\bf k}} \rangle  = 
\sum \limits_{i=N_1}^{N_2} |\Psi_{i{\bf k}}\rangle \sum \limits_{\bf q,q^{\prime}} c^{*}_{i,{\bf q}} ({\bf k}) a_{i,{\bf q^{\prime}}} ({\bf k})\langle {\bf k-q} | {\bf k-q^{\prime}} \rangle = \nonumber \\
& = \sum \limits_{i=N_1}^{N_2} |\Psi_{i{\bf k}}\rangle \sum \limits_{\bf q^{\prime}} c^{*}_{i,{\bf q^{\prime}}} ({\bf k}) a_{i,{\bf q^{\prime}}} ({\bf k})=  \sum \limits_{i=N_1}^{N_2} \tilde b_{i,n}({\bf k}) |\Psi_{i{\bf k}}\rangle = \nonumber \\
& =\sum \limits_{\bf q} \tilde \omega_{n,{\bf q}}({\bf k}) | {\bf k-q} \rangle, 
\label{Wannier:PW}
\end{eqnarray}
\begin{equation}
\tilde b_{i,n}({\bf k}) \equiv \sum \limits_{\bf q^{\prime}} c^{*}_{i,{\bf q^{\prime}}} ({\bf k}) a_{i,{\bf q^{\prime}}} ({\bf k}), \qquad
\tilde \omega_{n,{\bf q}}({\bf k}) \equiv \sum \limits_{i=N_1}^{N_2} \tilde b_{i,n}({\bf k}) c_{i,{\bf q}} ({\bf k}).
\end{equation}

To generate orthogonalized WFs, one should define the overlap matrix:
\begin{equation}
O_{nn^{\prime}}({\bf k}) \equiv \langle \widetilde{W}_{n{\bf k}} | \widetilde{W}_{n^{\prime}{\bf k}} \rangle= 
\sum \limits_{i=N_1}^{N_2} \tilde b_{i,n}^{*}({\bf k}) \tilde b_{i,n^{\prime}}({\bf k}).
\end{equation}

Orthogonalized Wannier functions are then obtained as:
\begin{equation}
|W^{\bf T}_n\rangle = \sum_{\bf k} |W_{n{\bf k}}\rangle e^{i{\bf kT}}, \qquad \mbox{where}
\end{equation}
\begin{eqnarray}
\label{WF-orth}
|W_{n {\bf k}}\rangle = \sum \limits_{n^{\prime}} (O_{nn^{\prime}}({\bf k}))^{-\frac{1}{2}} | \widetilde{W}_{n^{\prime}{\bf k}} \rangle = \sum \limits_{i=N_1}^{N_2} b_{i,n}({\bf k}) |\Psi_{i{\bf k}}\rangle =\sum \limits_{\bf q} \omega_{n,{\bf q}}({\bf k}) | {\bf k-q} \rangle 
\nonumber \\
b_{i,n}({\bf k}) \equiv \sum \limits_{n^{\prime}} (O_{nn^{\prime}}({\bf k}))^{-\frac{1}{2}}\tilde b_{i,n^{\prime}}({\bf k}) \qquad
\omega_{n,{\bf q}}({\bf k}) \equiv \sum \limits_{n^{\prime}} (O_{nn^{\prime}}({\bf k}))^{-\frac{1}{2}}\tilde\omega_{n^{\prime},{\bf q}}({\bf k}) .
\end{eqnarray}

For practical use---which includes generating the hamiltonian matrix for the kinetic energy term in the model Hamiltonian and performing constrained LDA/GGA calculation, one needs to compute the matrix elements of a given 
operator in the basis of the WFs. This can be conveniently done in reciprocal space. 

The matrix elements of the one-electron Hamiltonian in reciprocal space are defined as:
\begin{eqnarray}
\label{Wannier:Ham-k}
H^{WF}_{nm}({\bf k}) &= \langle W_{n{\bf k}} | \left (  \sum \limits_{i=N_1}^{N_2} |\Psi_{i{\bf k}} \rangle \varepsilon_i({\bf k})\langle \Psi_{i{\bf k}} | \right ) | W_{m{\bf k}} \rangle = \nonumber \\
& =  \sum \limits_{i=N_1}^{N_2} b^{*}_{i,n}({\bf k}) b_{i,m}({\bf k}) \varepsilon_i({\bf k}),
\end{eqnarray}
where $\varepsilon_i({\bf k})$ is the eigenvalue of the one-electron Hamiltonian for band $i$.

In real space, the Hamiltonian matrix reads:
\begin{eqnarray}
\label{Wannier:Ham}
H^{WF}_{nm}(\bf{T^{\prime}-T})&= \langle W^{\bf T}_n | \left ( \sum \limits_{\bf k} \sum \limits_{i=N_1}^{N_2} |\Psi_{i{\bf k}} \rangle \varepsilon_i({\bf k})\langle \Psi_{i{\bf k}} | \right ) | W^{\bf T^{\prime}}_m \rangle = \nonumber \\
& = \sum \limits_{\bf k} H^{WF}_{nm}({\bf k}) e^{i{\bf k(T^{\prime}-T)}}.
\end{eqnarray}

The Wannier functions occupancy matrix $Q^{WF}_{nm}$ is given by:
\begin{eqnarray}
\label{Wannier:Q}
Q^{WF}_{nm} &= \langle W^0_n | \left ( \sum \limits_{\bf k} \sum \limits_{i=N_1}^{N_2} |\Psi_{i{\bf k}} \rangle \theta(\varepsilon_i({\bf k}) - E_f ) ({\bf k})\langle \Psi_{i{\bf k}} | \right ) | W^0_m \rangle = \nonumber \\
& = \sum \limits_{\bf k} \sum \limits_{i=N_1}^{N_2} b^{*}_{i,n}({\bf k}) b_{i,m}({\bf k}) 
\theta(\varepsilon_i({\bf k}) - E_f ), \label{Wannier:Occ}
\end{eqnarray}
where $\theta$ is the step function and $E_f$ is the Fermi energy.

The transformation from plane waves to the Wannier functions basis is determined 
by Eqs. (\ref{WF-defin}-\ref{WF-orth}) and for the matrix elements by 
Eqs. (\ref{Wannier:Ham}) and (\ref{Wannier:Occ}). It can be convenient 
also to  back transform from WFs to the plane-waves basis. For example, 
if some constrained potential has a diagonal form in the WFs basis, 
$H_{nn^{\prime}} = \delta V_n\delta_{nn^{\prime}}$, it can be written 
in the plane-waves basis as: 

\begin{eqnarray}
\label{constr}
\widehat{H}_{constr}({\bf k}) & = & \sum \limits_n |W_{n{\bf k}} \rangle \delta V_n \langle W_{n{\bf k}} | \nonumber \\
H^{constr}_{qq^{\prime}}({\bf k}) & = & \langle {\bf k-q} | \widehat{H}_{constr}({\bf k}) | {\bf k-q^{\prime}} \rangle = \nonumber \\
& = & \sum \limits_n  \langle {\bf k-q}|W_{n{\bf k}} \rangle \delta V_n \langle W_{n{\bf k}} | {\bf k-q^{\prime}} \rangle = \nonumber \\
& = & \sum \limits_n \omega^{*}_{n,{\bf q}}({\bf k}) \delta V_n \omega_{n,{\bf q^{\prime}}}({\bf k}).
\end{eqnarray}

\subsection{Hamiltonian}
The localized Wannier functions are used as a basis set to define the Hamiltonian for strongly correlated 
materials:

\begin{eqnarray}
\label{Ham}
\widehat{H} & = & \sum \limits_{n m, \bf T \bf T^{\prime}} H^{WF}_{nm}({\bf T^{\prime}-\bf T}) \hat{c}^\dagger_{n\bf T}\hat{c}_{m\bf T^{\prime}} + \frac{1}{2}\sum \limits_{nm (n\ne m), \bf T}U_{nm}\hat{n}_{n\bf T}\hat{n}_{m\bf T} - \widehat{H}_{DC}
\end{eqnarray}
Here $\hat{c}^\dagger_{n\bf T} (\hat{c}_{n\bf T} )$ is the creation (annihilation) operator for an electron in the 
state defined by the Wannier function $|W^{\bf T}_n \rangle$, $\hat{n}_{n\bf T}$ is the occupancy operator for this state, and $\widehat{H}_{DC}$ is a correction term to avoid double-counting for the Coulomb interaction that is already taken into account in DFT in an averaged way. 

The first Hamiltonian term, on the right-hand side of Eq.(\ref{Ham}), corresponds to 
the kinetic energy, and the matrix elements $ H^{WF}_{nm}({\bf T^{\prime}-\bf T})$ are 
defined by Eq.(\ref{Wannier:Ham}). The Coulomb interaction is described by the second 
term on the right-hand side of Eq.(\ref{Ham}). The Coulomb matrix $U_{nm}$ can be nonzero, 
in the most general case, for all orbitals. However, normally one takes into account only 
interactions between electrons in WFs having the symmetry of the transition-metal-ion 
$d$ or $f$ orbitals. The corresponding matrix elements can be obtained  using only two 
parameters: the direct Coulomb interaction parameter $U$ and the exchange Coulomb 
interaction (Hund) parameter $J$ (see for details Ref.~\onlinecite{ani97}). 
The double-counting correction is taken as:
\begin{eqnarray}
\label{Ham-DC}
\widehat{H}_{DC} & = &  \sum \limits_{n \bf T}\epsilon_{DC} \hat{n}_{n\bf T}.   
\end{eqnarray}
Here we focus on a $d$-electron system and use for the double-counting potential:
\begin{eqnarray}
\label{Pot-DC}
\epsilon_{DC}=U(n_d-\frac{1}{2}), 
\end{eqnarray}
with $n_d$ the number of $d$ electrons\cite{d-electrons}. Eqs. (\ref{Ham-DC}-\ref{Pot-DC}) correspond to the   
assumption that the contribution to the DFT total energy from the Coulomb interaction 
between $d$ electrons is given by: 
\begin{eqnarray}
\label{E-DFT}
E_{DFT}=\frac{1}{2}Un_d(n_d-1)
\end{eqnarray}
and the fact that one-electron energies in DFT are derivatives of the total energy with respect to the orbitals occupancy
\begin{eqnarray}
\label{eps-DFT} 
\epsilon_d=\frac{\partial E}{\partial n_d}.
\end{eqnarray}

\subsection{Constrained DFT calculations for the Coulomb interaction parameter $U$}

In a rigorous way, the Coulomb interaction parameter $U$ for electrons in a state
described by Wannier function $W_n({\bf r})=\langle{\bf r}|W^{0}_n \rangle$ should be calculated via the integral:
\begin{eqnarray}
\label{integral}
U=\int d^3r d^3r'|W_n({\bf r})|^2U({\bf r},{\bf r}')|W_n(\bf r')|^2, 
\end{eqnarray}
where the screened Coulomb interaction $U({\bf r},{\bf r}')$ is defined via the operator equation:
\begin{eqnarray}
\label{Uscr}
U=[1-vP]^{-1}v
\end{eqnarray} 
with the bare Coulomb interaction, $v({\bf r},{\bf r}')=1/({\bf r}-{\bf r}')$, and the polarization operator $P$:
\begin{eqnarray}
\label{P}
P({\bf r},{\bf r}')=\sum^{occ}_{i}\sum^{unocc}_{j}\psi_i({\bf r})\psi^*_i({\bf r}')\psi^*_j({\bf r})\psi_j({\bf r}')\left\{ \frac{1}{\epsilon_i-\epsilon_j+i0^+} - \frac{1}{\epsilon_j-\epsilon_i-i0^+}\right\}.
\end{eqnarray} 
However various attempts\cite{Solovyev,Ferdi} to use Eqs. (\ref{integral}-\ref{P}) to calculate the Coulomb interaction parameter $U$ gave large variations in the resulting values. This is due to the many possibilities in choosing the channels of the screening via the definition of a set of occupied and unoccupied states in Eq.(\ref{P}). It either gives a strong underestimation of the $U$ values, when one includes into the screening channels transitions between $d$-states (that is clearly unphysical because electrons cannot screen themselves), or too large values of $U$,  when only states other than $d$-states are included in the sums in Eq.(\ref{P}), but the summation over higher and lower energy states is not extended far enough.

Another way to determine the Coulomb interaction parameter $U$ is to use constrained DFT 
calculations,\cite{AG,ConstrainPickett,Coco,nakamura} where the screening and relaxation effects 
are taken into account explicitly in a self-consistent procedure. If one uses the   
assumption that the contribution to the DFT total energy from the interaction 
between $d$-electrons is given by Eq.(\ref{E-DFT}), then $U$ can be calculated as a second 
derivative of the DFT total energy with respect to the $d$-orbital occupancy:
\begin{eqnarray}
U=\frac{\partial^2 E_{DFT}}{\partial^2 n_d}, 
\end{eqnarray}
using Eq.(\ref{eps-DFT}) this can be expressed via the first derivative of the one-electron energy $\epsilon_d$:
\begin{eqnarray}
\label{U-constr} 
U=\frac{\partial \epsilon_d}{\partial n_d}.
\end{eqnarray}
To use Eq.(\ref{U-constr}), one needs to perform DFT calculations with a constraint fixing the 
occupancy of $d$-orbitals  to certain values. In practice, this is done via an auxiliary 
potential in the form of a projection operator acting on $d$-symmetry WFs $|W_{n}^T \rangle$:
\begin{eqnarray}
\label{dV}
\widehat{H}_{constr} & = &  \sum_n |W_{n}^T \rangle \delta V_n \langle W_{n}^T | 
\end{eqnarray}
(in reciprocal space this equation will take the form of Eq.(\ref{constr})). One-electron energy can be calculated as diagonal elements of the Hamiltonian matrix, in Eq.(\ref{Wannier:Ham}), and the corresponding occupancy as diagonal elements of the occupation matrix, in Eq.(\ref{Wannier:Q}):
\begin{eqnarray}
\label{e-Q}
\epsilon_d & = & H^{WF}_{dd}(0) \\ \nonumber
n_d & = & Q^{WF}_{dd},
\end{eqnarray}
and the derivative $\frac{\partial \epsilon_d}{\partial n_d}$ (Eq.(\ref{U-constr})) is computed then numerically.


\subsection{Dynamical Mean-Field Theory}
The problem defined by the Hamiltonian (\ref{Ham}) can be
solved by any of the methods developed to treat many-body effects.
In the present work we have used Dynamical Mean-Field 
Theory (DMFT)~\cite{vollha93,pruschke,georges96} which was recently found to be a
powerful tool to numerically solve multiband Hubbard models. 

In DMFT, the lattice problem becomes an effective single-site
problem, which has to be solved self-consistently
for the matrix self-energy ${\widehat{ \Sigma}} $ and the local
matrix Green function in the WFs basis set:
\begin{eqnarray}
\label{gloc} G_{nn'}(\varepsilon)=\!\frac{1}{V_{BZ}} \int d{\bf k}
\left( \left[ \;(\varepsilon-\mu)
\widehat{1}-\widehat{H}^{0}({\bf k})
-\widehat{\Sigma}(\varepsilon)\right]^{-1}\right)_{nn'},
\end{eqnarray}
where $\mu$ is the chemical potential, $\widehat{H}^{0}$ is the noninteracting part of the Hamiltonian 
(\ref{Ham}): 
\begin{eqnarray}
\label{Ham0}
\widehat{H}^0 & = & \sum \limits_{n m, \bf T \bf T^{\prime}} H^{WF}_{nm}({\bf T^{\prime}-\bf T}) \hat{c}^\dagger_{n\bf T}\hat{c}_{m\bf T^{\prime}} - \widehat{H}_{DC}
\end{eqnarray}
(in reciprocal space, the matrix elements for $\widehat{H}^{0}({\bf k})$ can be calculated using 
Eq.(\ref{Wannier:Ham-k}))
and $\widehat{\Sigma}(\varepsilon)$ is the self-energy in the Wannier functions basis:
\begin{eqnarray}
\label{Sigma}
  \widehat{\Sigma}(\varepsilon)= \sum_{nn'} |W_{n} \rangle \Sigma_{nn'}(\varepsilon) \langle W_{n'}|.
\end{eqnarray}

The DMFT single-site problem may be viewed as a self-consistent single-impurity 
Anderson model~\cite{georges96}. The corresponding
local one-particle matrix Green function ${\widehat{G}}$ can be
written as a functional integral~\cite{georges96} involving an
action where the Hamiltonian of the correlation problem under
investigation, including the interaction term with the Hubbard
interaction, enters~\cite{LDADMFT}. The
action depends on the bath matrix Green function ${\widehat{\cal
{G}}}$ through
\begin{eqnarray}
\label{G0} {(\widehat{\cal {G}})}^{-1} = (\widehat{G})^{-1} +
{\widehat{\Sigma}}.
\end{eqnarray}
To solve the functional integral of the effective single-impurity Anderson
problem, various methods can be used: quantum Monte Carlo (QMC),
numerical renormalization group (NRG), exact diagonalization (ED),
noncrossing approximation (NCA), etc. (for a brief overview of the methods 
see Ref.~\onlinecite{LDADMFT}). In the present work, the QMC \cite{Hirsh}
method is used to solve the impurity problem. The  real-frequencies  
single-particle spectral functions are computed using the maximum entropy 
method.\cite{Jarrell96}

\section{Calculations for NiO}
\label{Cdetails}
The calculation scheme described in Section \ref{Method} was applied in the 
present work to the problem of the classical Mott insulator nickel oxide, NiO.
A DMFT study of the NiO electronic structure was done recently\cite{Kunes07} with LMTO-based Wannier
functions and here we use previous work as a reference.
Actually, in the Zaanen-Sawatzky-Allen classification,\cite{ZSA} NiO
is normally considered as a 'charge transfer insulator' where the minimal energy excitation across 
the gap happens between occupied oxygen $p$-states and unoccupied transition 
metal $d$-states. This fact makes it crucial to include in the calculation not 
only partially filled $d$-states (as is usually done for strongly correlated 
materials), but also explicitly take into account oxygen $p$-orbitals.

The pseudopotential plane-waves method, as implemented in the Quantum-ESPRESSO \cite{ESPRESSO} package, was 
used in the present work. 
The calculations were performed within the local density approximation to density-functional theory, using
Vanderbilt ultrasoft pseudopotential \cite{Vanderbilt} in the Rappe-Rabe-Kaxiras-Joannopoulos form.\cite{RRKJ} For
the lattice parameter of NiO (rock-salt structure), we used the experimental value $a=4.193$ \AA. A kinetic-energy cutoff of 40 Ry was employed for the plane-wave expansion of the electronic states. 
The integrations in reciprocal space were performed using a (4,4,4) Monkhorst-Pack \cite{MonkhorstPack} k-point grid.

The Wannier functions, introduced in Section \ref{Method}, are defined by the 
choice of the Bloch functions Hilbert space and by a set of trial localized 
orbitals. To show how this choice influences the results, we performed two 
calculations for the WFs. One with only bands formed by the Ni $3d$ states 
included in the Hilbert functional space and the second one where this space 
was extended by inclusion also of oxygen $p$-bands. In both cases, the 
projection was done on pseudoatomic wave functions.

\begin{figure}[t]
\begin{center}
\epsfxsize=8cm
\epsfbox{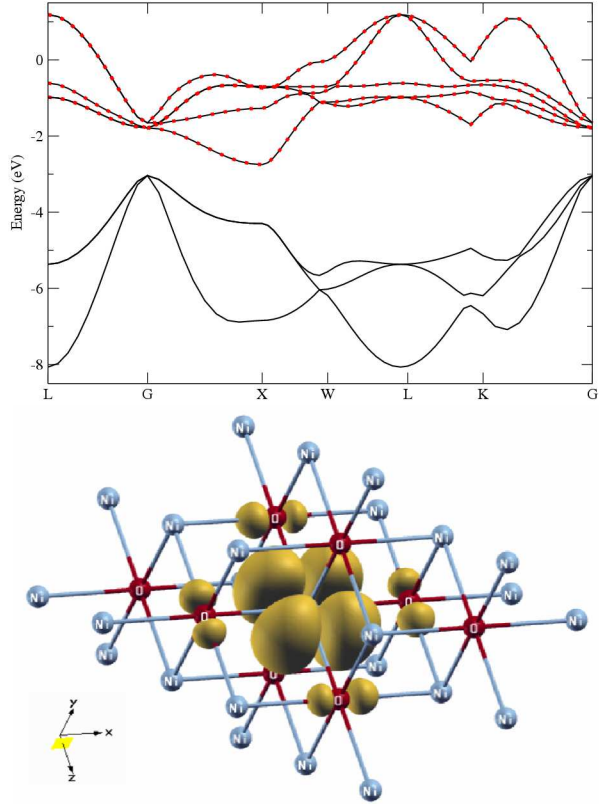}
\end{center}
\caption{(Color online) {\bf Top frame:} Calculated LDA band structure of NiO (solid line) and bands obtained from the five $d$-type Wannier functions
 (circles). The energy window used to compute the Wannier functions was set
to [-2.5,+1.5] eV from the Fermi level. {\bf Bottom frame:} Charge density of the corresponding $d_{xz}$-like Wannier function (plotted using the XCrysDen\cite{xcrys}
package).}
\label{fig:dbands}
\end{figure}

\begin{figure}[t]
\begin{center}
\epsfxsize=8cm
\epsfbox{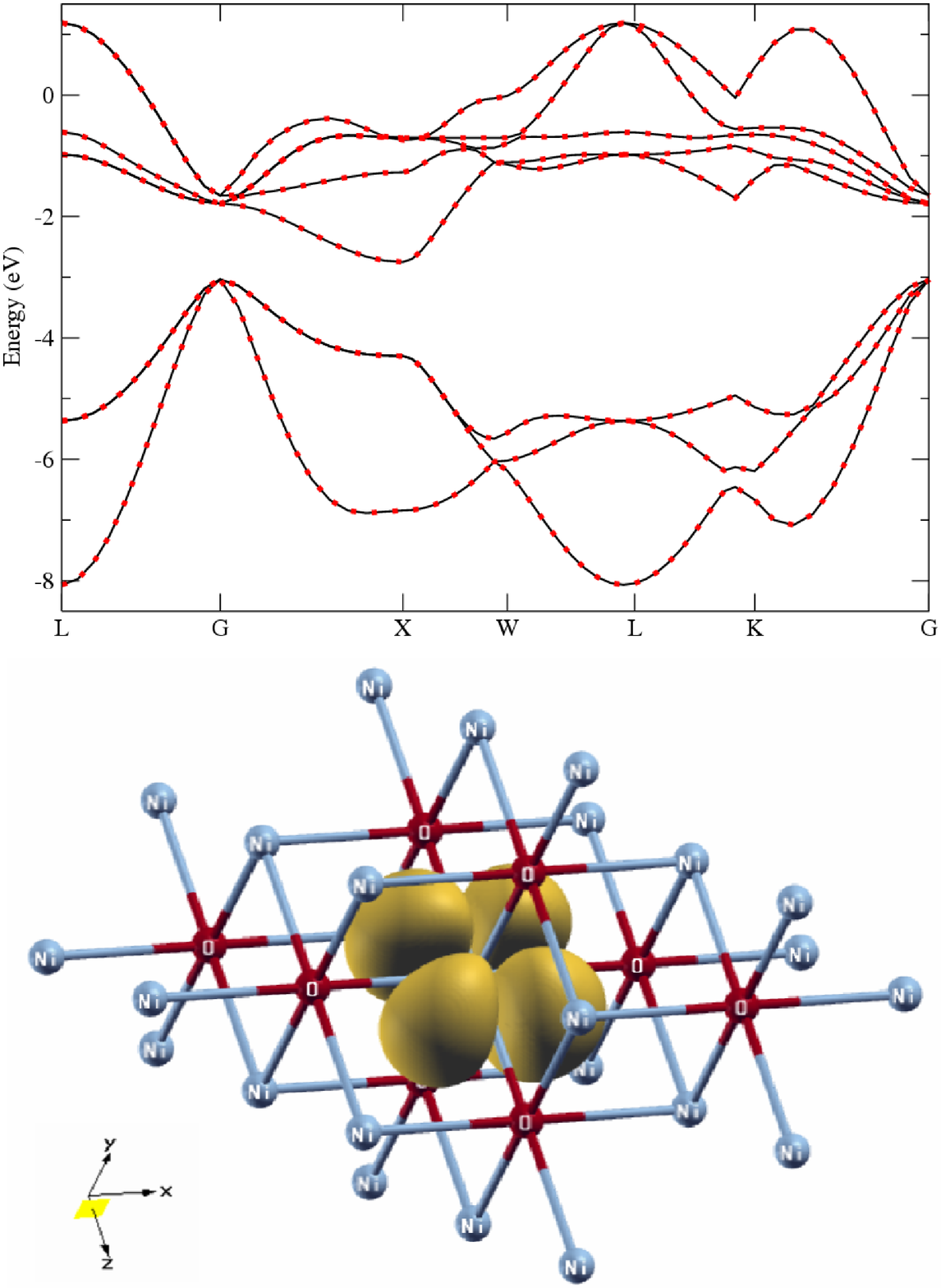}
\end{center}
\caption{(Color online) {\bf Top frame:} Calculated LDA band structure of NiO  (solid line) and bands obtained from the eight Wannier functions (circles).
The energy window used to compute the Wannier functions was set to [-8.5,+1.5] eV from the Fermi level. {\bf Bottom frame:} Charge density of the corresponding  $d_{xz}$-like Wannier function.}
\label{fig:allbands}
\end{figure}

The results obtained without and with the $p$-bands are presented in Fig.\ref{fig:dbands} 
and Fig.\ref{fig:allbands}, respectively. In both cases, the $d$-bands obtained in the 
pseudopotential calculations are exactly reproduced by the bands obtained diagonalizing 
the k-space Hamiltonian matrix (Eq.(\ref{Wannier:Ham-k})). However the spatial distributions 
of the corresponding Wannier functions are different in the two cases. To demonstrate that, spatial distribution of $d_{xz}$-like Wannier function charge density was plotted (see Fig.\ref{fig:dbands} and Fig.\ref{fig:allbands}). Omission of the 
oxygen bands in the first calculation results in the appearance of a contribution from 
$p$ orbitals on neighboring oxygen atoms in the Wannier function obtained by projection 
of the Ni-atom $3d$ orbitals (see Fig.\ref{fig:dbands}). Extending the Hilbert space with 
explicit inclusion of oxygen $p$ bands results in a Wannier function that is nearly a 
pure $d$-orbital (see Fig.\ref{fig:allbands}). In the following, we will use WF obtained 
with the full sets of bands and pseudo-atomic orbitals including all Ni $3d$ and O $2p$ states.

\begin{figure}[t]
\begin{center}
\epsfxsize=10cm
\epsfbox{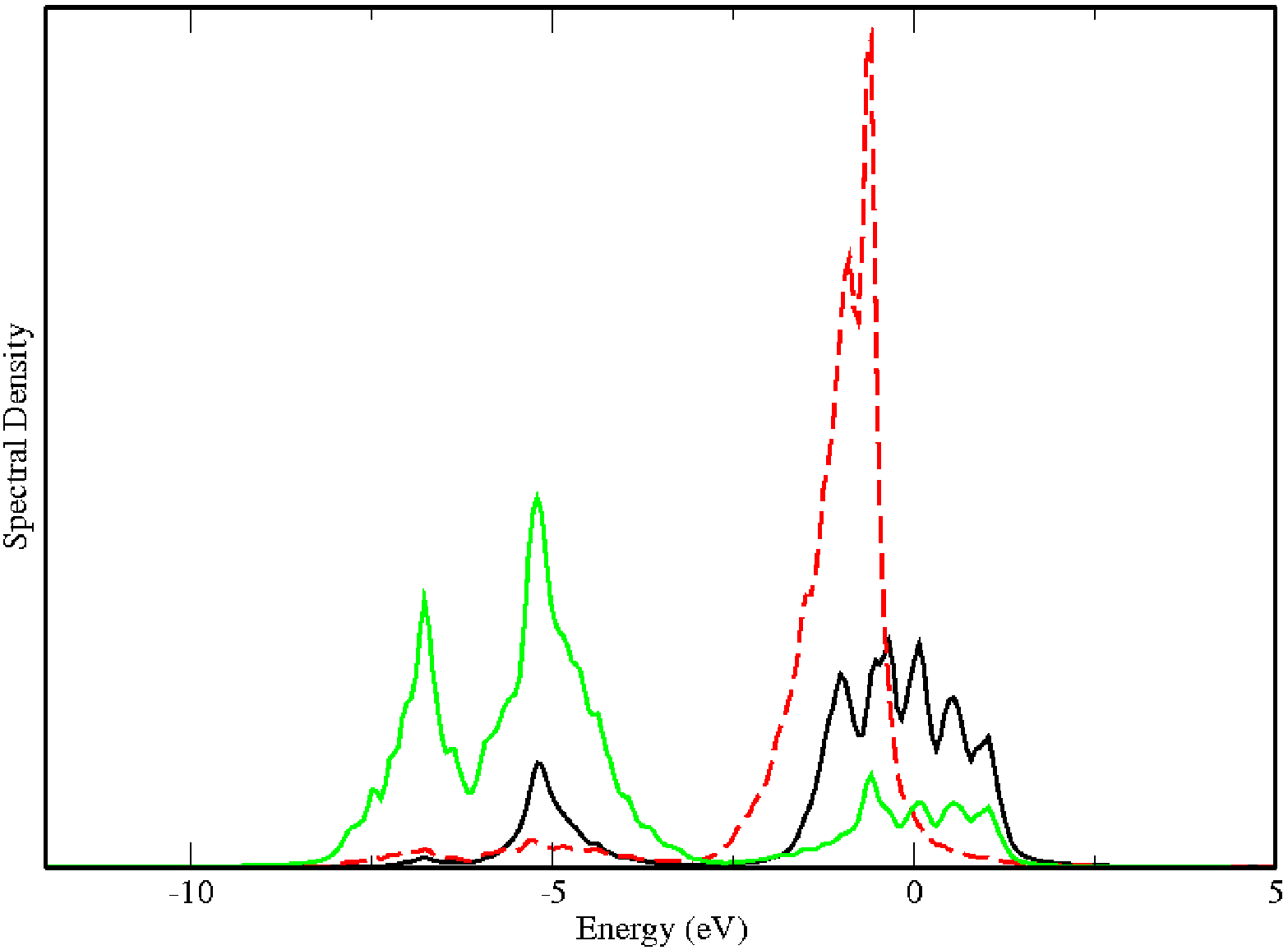}
\end{center}
\caption{(Color online) Orbital resolved density of states for the non-interacting LDA hamiltonian. Solid black line: $e_g$ states of Ni, dashed red line: $t_{2g}$ states of Ni, solid green line: p states of O. Here and later Fermi energy
corresponds to zero.}
\label{fig:ldados}
\end{figure}

The partial densities of states calculated from the WF hamiltonian are shown in 
Fig.\ref{fig:ldados}. One can distinguish the occupied oxygen $p$-band and, at 
higher energy, the partially occupied $d$-band. The $t_{2g}$ 
sub-band is nearly filled, while the Fermi level is located essentially in the 
middle of the $e_g$ sub-band. 

\begin{figure}

\begin{center}
\epsfxsize=10cm
\epsfbox{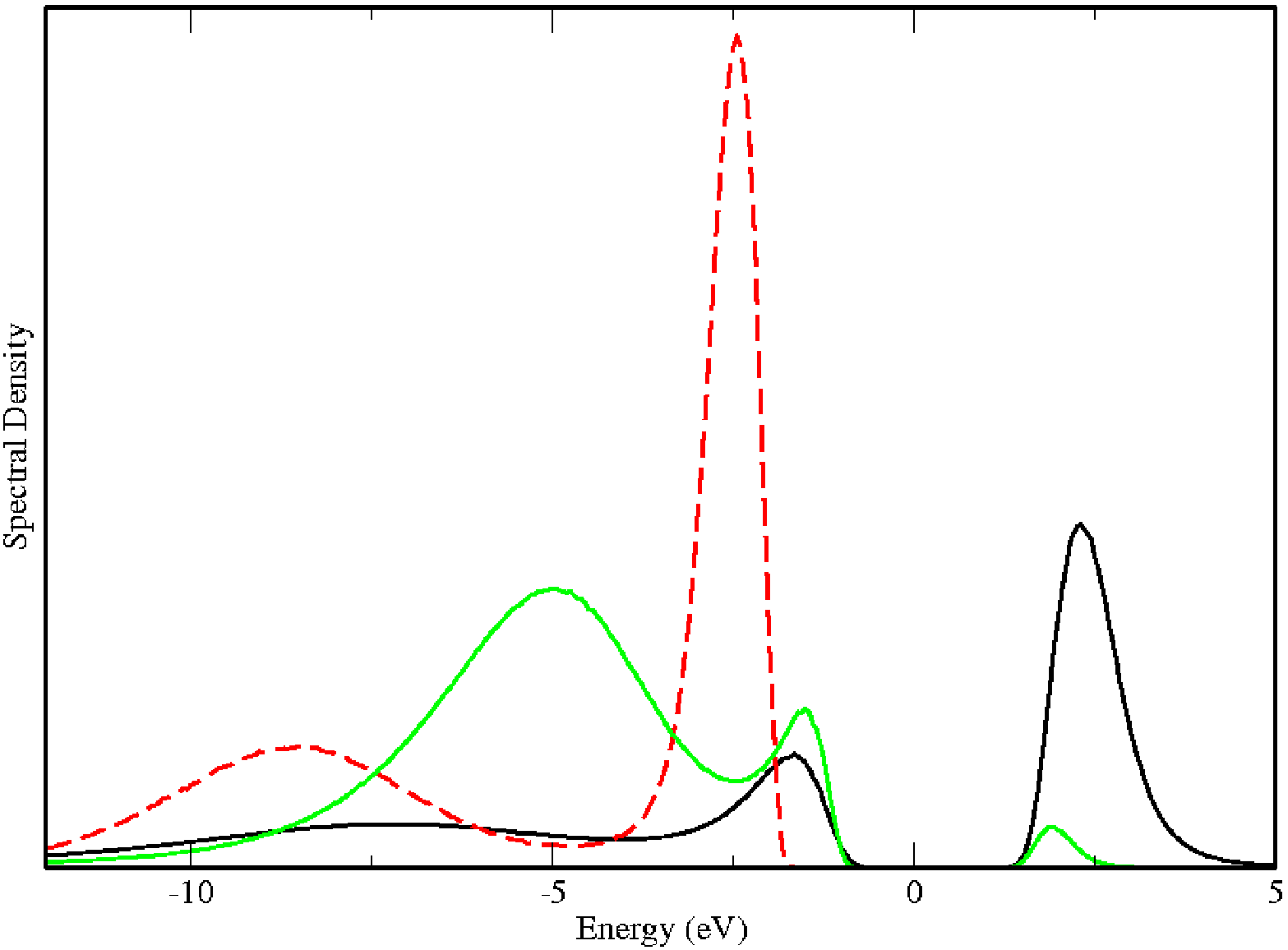}
\end{center}
\caption{
(Color online) Orbital resolved density of states of NiO for the interacting 
Hamiltonian. Solid black line: $e_g$ states of Ni, dashed red line: $t_{2g}$ 
states of Ni, solid green line: $p$-states of O.}

\label{fig:dmftdos}
\end{figure}

\begin{figure}[t]
\begin{center}
\epsfxsize=10cm
\epsfbox{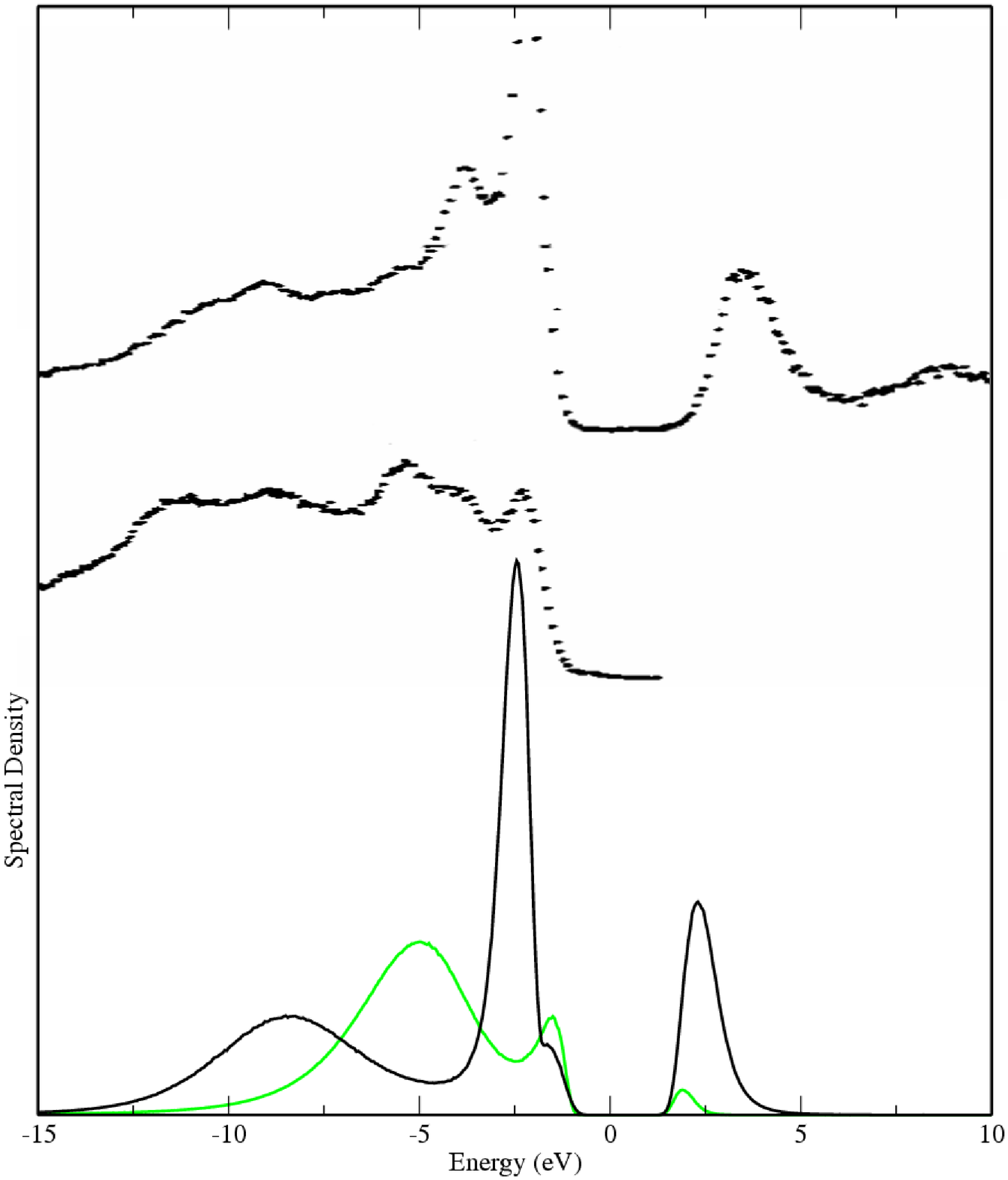}
\end{center}
\caption{(Color online) Theoretical (solid lines) spectral density of NiO for the
interacting Hamiltonian, compared to photoemission (direct and inverse) data (dotted line) obtained at 120
eV (top curve) and 66 eV (second-top curve) photon energies (Ref.~\onlinecite{saw84}). Solid black line: Ni $3d$ states. Solid green line: $2p$-states of O.}
\label{fig:DOS}
\end{figure}
The calculation of Coulomb interaction parameter $U$ was done using straightforward
constrained DFT procedure\cite{Ferdi}. 
We performed a self-consistent calculation in a supercell containing two 
formula units. The constrained potential $\delta V_n$ (Eq.(\ref{dV})) on the first Ni 
atom was positive and equal to 0.1 Ry, and for the second Ni atom it had the same 
magnitude and opposite sign. The Wannier functions occupations and energies were 
calculated and the $U$ value was then evaluated as described in Sec.~\ref{Method}, 
using Eqs. (\ref{U-constr}-\ref{e-Q}). The resulting value for the Coulomb interaction 
parameter $U$ was found to be 6.6 eV. While this is smaller than the value of 8 eV 
obtained from constrained LDA calculations for linear-muffin-tin orbitals,\cite{AZA} 
it is larger than the value of  of 4.5 eV obtained from linear-response-theory 
calculations for pseudo-atomic orbitals.\cite{Coco}  

The interacting Hamiltonian, constructed from the pseudopotential plane-waves 
calculations, was solved by DMFT-QMC simulations. In the QMC simulations, the 
inverse temperature value was $\beta=10$ eV$^{-1}$ ($T = 1160 K$), and we 
used 80 time slices and $4 \times 10^6$ Monte Carlo sweeps. Other DMFT+QMC-calculation
details are the same as in \cite{Kunes07}.
The DMFT calculations result in strong changes in the spectral functions 
(see Fig.\ref{fig:dmftdos}) compared to the LDA densities of states 
(Fig.\ref{fig:ldados}), consistent with recent DMFT calculations based on 
the LMTO method.\cite{Ren06,Kunes07} First of all, the correlated spectra 
show a large-gap insulator in contrast to the metallic LDA solution. 
The half filled $e_g$ band is now split into an empty upper-Hubbard band 
and an occupied lower-Hubbard band.  In addition, with explicit inclusion 
of the oxygen $2p$ band, the lower Hubbard $e_g$ band is found to largely 
overlap in energy with the oxygen $p$-band and to strongly hybridize with it. 
Furthermore, although the $t_{2g}$ band remains occupied, it is shifted down 
in energy so that it also overlaps with the oxygen band. 

In Fig.\ref{fig:dmftdos}, we present a comparison of the experimental 
photoemission (direct and inverse) spectra\cite{saw84} with the spectral 
functions obtained from our DMFT pseudopotential calculations. 
The photoemission spectra measured with 120 eV photons reflect more $3d$ 
Ni states and those with 66 eV have more oxygen contributions, due to the 
difference in the cross section values. One can observe a good agreement 
between the experimental and calculated spectra. 
The calculations reproduce a large band gap as well as remarkable 
redistribution of spectral weight from the top of the valence band to 
lower energies, going from the 120 eV to the 66 eV spectrum.
%
%
Comparison 
with the theoretical orbital-resolved result in Fig.\ref{fig:dmftdos}, 
indicate that this is consistent with a decreasing $d$-states spectral 
density contribution and an increasing $p$-states contribution to the spectrum.

\section{Conclusion}
We have presented a computational scheme for strongly correlated materials 
where, in the framework of the pseudopotential plane-waves method, 
atomic-centered Wannier functions are calculated and the Hamiltonian matrix 
elements are evaluated in the Wannier-functions basis. Then, via constrain 
DFT calculations, the Coulomb interaction parameter $U$ is obtained for 
electrons in these Wannier functions, allowing us to generate in a consistent 
way all necessary  components of the Hamiltonian. This Hamiltonian is solved 
by Dynamical-Mean-Field Theory with the numerically exact Quantum Monte-Carlo 
method as effective-impurity-problem solver. We have applied this scheme to 
nickel oxide and demonstrate good agreement with the experimental photoemission data.

\section{Acknowledgments}
This work was supported by the Russian Foundation for Basic Research under 
the grant RFFI 07-02-00041. We are grateful to 
Jan Kunes and Dieter Vollhardt for helpful discussions and for supplying 
the QMC-DMFT computer code. We also acknowledge support for this work by 
the Light Source Theory Network, LighTnet, of the EU.


\begin{thebibliography}{100}
\bibitem{d-electrons}
By $d$ electrons we imply electrons in Wannier orbitals of $d$ symmetry.
\end{thebibliography}
\end {document}